\documentclass[useAMS,usenatbib,referee]{article}
\usepackage{amsfonts}
\usepackage{amsmath}
\usepackage{amssymb}

\usepackage[T1]{fontenc}

\usepackage[dvips]{graphicx}
\usepackage{myaasmacros}
\usepackage{color}
\usepackage{subfig}
\usepackage{multirow}
\usepackage{ulem}
\def\bea{\begin{eqnarray}}
\def\ena{\end{eqnarray}}

\newcommand{\mr}[1]{\mathrm{#1}}

\begin{document}

\title{Irregularities in the rate of generation of giant pulses  from the  Crab pulsar  observed at 111 MHz}

\author {A.~N.~Kazantsev$^{1}$\thanks{E-mail:kaz.prao@bk.ru (ANK)},~V.~A.~Potapov$^{1}$\thanks{E-mail:potap@prao.ru (VAP)},~M.~S.~Pshirkov$^{2,3,1}$\thanks{E-mail:pshirkov@sai.msu.ru(MSP)}
~B.~Ya.~Losovskii$^{1}$\thanks{E-mail:blos@prao.ru (BYL)}
 ,\\
 $^{1}$P. N. Lebedev Physical Institute of the Russian Academy of Sciences,\\ Pushchino Radio Astronomy Observatory,
Pushchino 142290, Russia
\\
 $^{2}$
 Sternberg Astronomical Institute, Moscow State University, \\119992, 
 Universitetskiy Prospekt, 13, Moscow, Russia
 \\
 $^{3}$ Institute for Nuclear Research of the Russian Academy of Sciences,\\ 117312, Moscow, Russia
}

\date{}


\maketitle

\label{firstpage}

\begin{abstract}

We present analysis of the rate of giant radio pulses (GPs) emission from the Crab pulsar (B0531+21). Results of our 9 years daily observations with the Large Phased Array radio telescope of Pushchino Radio Astronomy Observatory at 111 MHz were used. Limited sample of 8753 strong individual pulses in 2004 observational sessions was further analysed. It was shown that the observed monthly averaged rate of GPs emission was highly unstable during the entire span of observations and changes by about two orders of magnitude for high-energy pulses. Data were further analysed to search for the possible connection between pulsar glitches and the process of GP emission. We have found a significant increase in the rate of emission of high-energy GPs after MJD 58064, when the largest glitch ever observed in the Crab pulsar was happened. Although considerable changes in GPs emission rate could have been caused by the propagation effects in the nebula itself, we have found that the pulsar had demonstrated high degree of intrinsic irregularity of high-energy pulses emission over long time intervals.

\end{abstract}


\section{Introduction}
\label{sec:intro}
Averaged profiles obtained from the thousands or even millions of individual pulses are highly stable for the majority of pulsars. However,  their individual pulses  could vary very much in shape, duration and strength due to fluctuations in the process of  pulse emission in pulsar's magnetosphere and to the effects arising during propagation of signal through the interstellar medium. The peak flux density and fluence of individual pulses can change in extremely wide limits. These strong fluctuation of flux was already noticed  in pioneer works about the Crab pulsar \cite{Staelin1968}. Soon it was found that this pulsar regularly emits  strong individual pulses with flux densities which  exceed average level by at least order of magnitude \cite{Sutton1971}.  Subsequently these pulses were dubbed as Giant  Pulses (GPs). 

The GPs are one of  the most intriguing phenomena in pulsar astrophysics. The brightest GPs of the Crab pulsar observed to date had flux density larger than 2 MJy at 9.25 GHz and width less than  0.4 ns \cite{Hankins2007}. The Crab pulsar is a known prolific source of regularly generated GPs which allows to investigate their characteristics in fine details \cite{Jessner2005,Popov2006,Mickaliger2012}. 

In order to build comprehensive set of strong GPs  and investigate their properties, especially the evolution of the production rate, we used almost 9 years (2010 Feb -- 2018 Nov, MJD 55240 -- 58440) of  daily observations of the Crab pulsar at frequency of 111 MHz at the Large  Phased Array radio telescope  of the Pushchino Radioastronomy Observatory of P.N.Lebedev Physical Institute (LPA LPI).   We were particularly interested in possible correlation of the GPs intensity and production rate with   glitching activity which is another well-known property of this young pulsar. The glitch process is a  rapid increase of rotational frequency  which is followed by gradual decay and almost complete recovery of pre-glitch values of frequency and its time derivative \footnote{http://www.jb.man.ac.uk/pulsar/glitches/gTable.html}.  Most likely these phenomena are caused by  the processes taking place inside the neutron star itself, such as unpinning of superfluid vortices from the crust \cite{Haskell2015}. The largest glitch in almost 50 years of observations  with a relative increase in frequency $\delta \nu/\nu=5.2\times10^{-7}$ occurred around 2017 November 08 (MJD=58064) and although no changes  in the pulse profile shape  were observed near the glitch epoch \cite{Shaw2018}, we used the possibility to check for possible changes in the GP properties. Without specifying any  certain theoretical model we wanted to check whether the  catastrophic glitch event  might cause some secondary  long-lasting perturbations in the magnetosphere structure, which would subsequently lead to a significant change in the process of generation of individual pulses and, particularly, GPs.

The paper is organized as follows: in  Section~\ref{sec:data} we describe the data and the method of data analysis, Section~\ref{sec:dis} contains our results and discussion, and we draw conclusions in Section~\ref{sec:conclusion}.

\section{Data and methods}
\label{sec:data}

The observations were carried out in 2010--2019  using the first  beam of the LPA LPI  transit radio telescope. 
The effective area of the telescope varied during observations from $20000\pm1300~\mathrm{m^2}$ (2010 - 2017.5 data) to $35000\pm2000~\mathrm{m^2}$ (2017.5 - 2018 data).  A  digital receiver with 512 channels  ($\delta f = 5$~kHz in each frequency channel) was used. The central frequency of observations was 111 MHz, the effective bandwidth of observations was about 2.3 MHz (460 channels were used).  Observation session was equal to the duration of passage of the Crab nebula across half-maximum of the antenna beam  -- 198 sec.   The sampling frequency was 2.4576 ms that was a good compromise between the radiometric gain and the time resolution of scattered pulsar profile. A single linear polarization was used. Totally, 3040 observational sessions have been conducted.

The data reduction was made in several steps. First, the effect of the frequency dispersion of pulse in the interstellar medium was  removed  off-line("dedispersion"). After this stage,  1036 sessions of observation were excluded from further consideration  due to strong ionospheric effect and high level of noise. At the next step, in each session we corrected the data for  the geometrical beam factor. Usually the antenna response for the beam factors is proportional to $\mathrm{Sinc}^2(t)$ function,  but for a substantial part of our data the influence of ionospheric effects was significant, leading to significant distortions of the theoretical antenna response. We used  more complex function to correct for these distortions. 

\begin{equation}
 f(x) = A \times\mathrm{Sinc}^2(t + t_{0}) + C_{0},
 \label{eq_beam:ref}
\end{equation}

where $A$  is the maximal value of beam function, $t_{0}$ -- shift of beam function along the temporal axis ,  $C_{0}$  is 'zero-level' -- the amplitude of background  in the direction of  Crab nebula. For several observational session  already mentioned ionospheric effects made full correction of the geometrical beam factor impossible. For that very reason, we used fitting by  13-order  polynomial to reduce remaining distortions.

At this stage we introduce absolute flux normalization using the Crab nebula as a calibrating source --  the baseline level was set at 1720 Jy \cite{Bietenholz1997,Smirnova2009}. That allowed us  to convert  flux density in analog digital converter units (ADC) to  Janskys  using  calibration coefficient:

\begin{equation}
 K [\frac{\mathrm{ADC units}}{\mathrm{Jy}}] = \frac{1720}{A - C_{0}},
 \label{calib_coeff:ref}
\end{equation}

Finally, we searched all sessions for giant pulses using methods of machine learning (see Appendix \ref{app-ML}).

\begin{figure}
\begin{center}
\includegraphics[scale=0.7, angle=0]{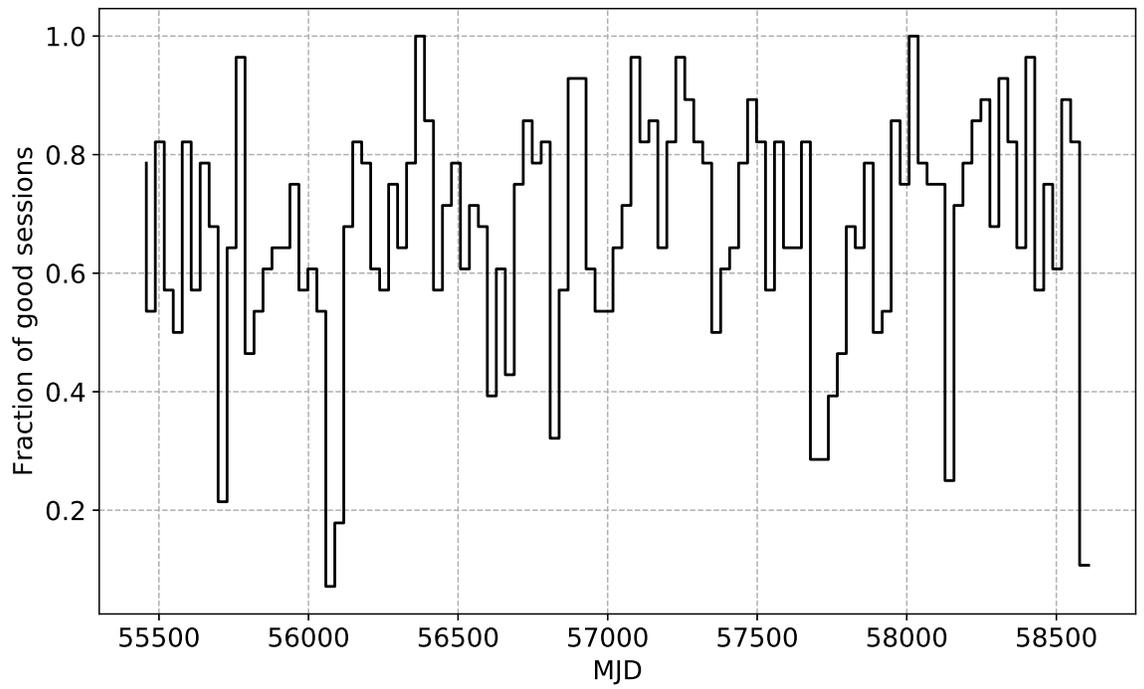}
\end{center}
\caption{Fraction  of good quality observational sessions. Bins are  30-days long.}
\label{fig:sessions}
\end{figure}

The total amount of detected pulses was equal to 8753 in 2004 good quality observational sessions which were performed during our time span of observations (Fig. \ref{fig:sessions}).

\section{Results and Discussion}
\label{sec:dis}

\begin{figure}
\begin{center}
\includegraphics[scale=1, angle=0]{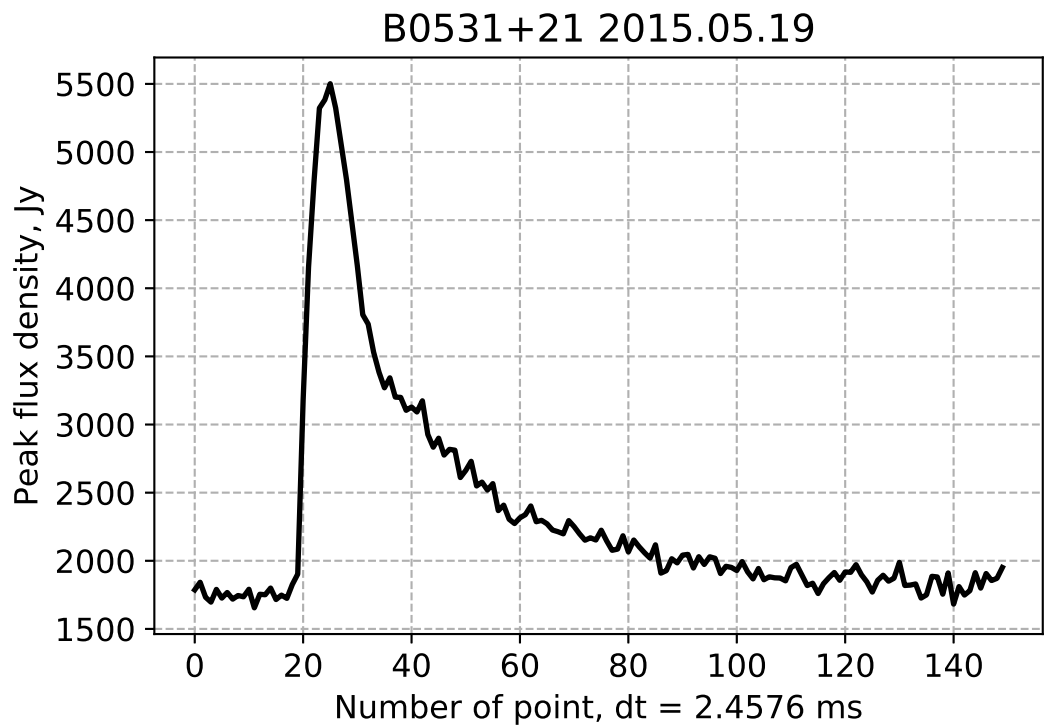}
\end{center}
\caption{An example GP with peak flux density equal to 3780 Jy and   broadening of pulse due to interstellar scattering  $\tau_d \approx 98$~ms.}
\label{fig:GP}
\end{figure}

\begin{figure}
\begin{center}
\includegraphics[scale=1, angle=0]{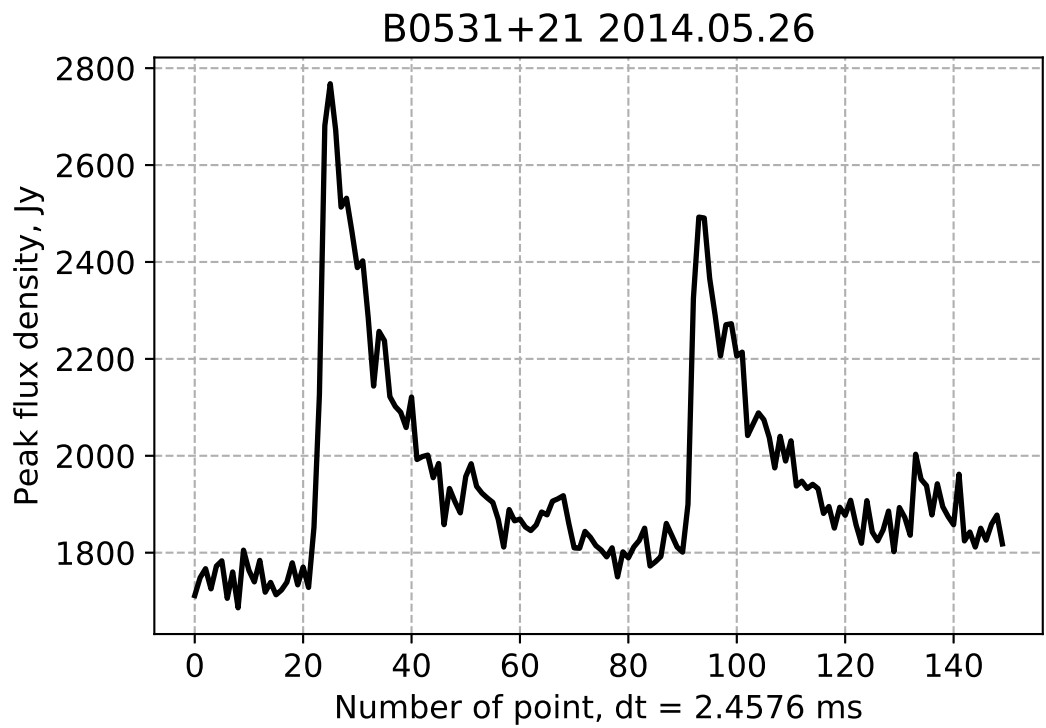}
\end{center}
\caption{An example of  sequence of two strong GPs in one session separated by 5 periods.}
\label{fig:double_GP}
\end{figure}

\begin{figure}
\begin{center}
\includegraphics[scale=0.7]{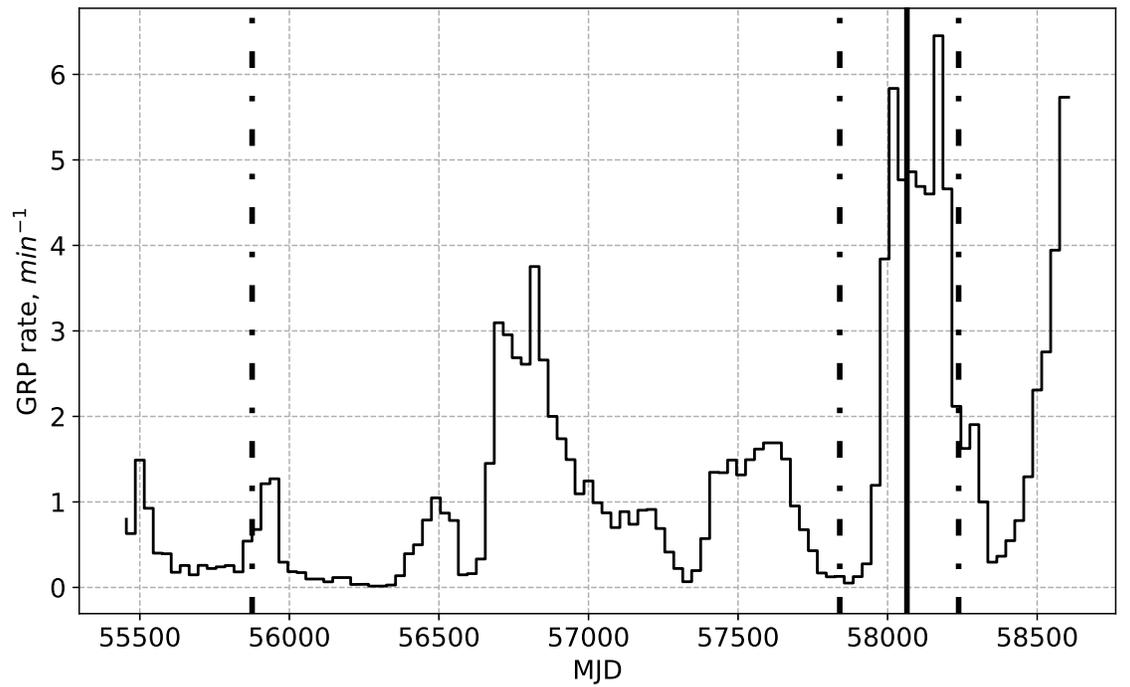}
\end{center}
\caption{The rate of GPs with peak flux $S$ larger than $4\sigma_{noise}$. Positions of glitches at MJD 55875.5, 57839.92, and 58237.357 are shown by black dash-dotted vertical lines. By far the strongest glitch at MJD 58064.55 is shown by a vertical bold  line.  Bin size is equal to 30  days and the rate is corrected for average number of good observational sessions in respective bins (see Fig. \ref{fig:sessions})}
\label{fig:rate}
\end{figure}

\begin{figure}
\begin{center}
\includegraphics[scale=0.6]{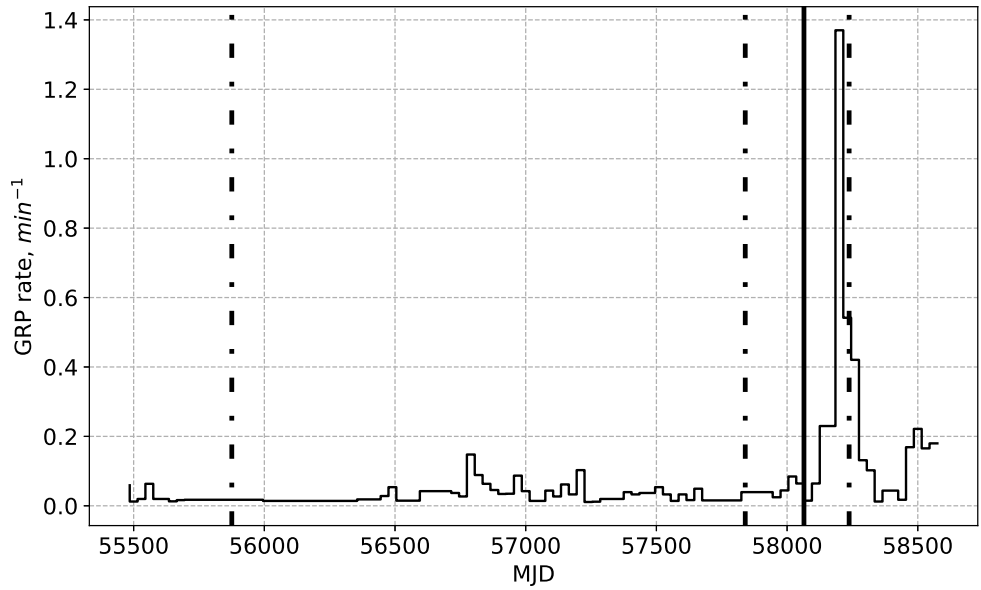}
\end{center}
\caption{The rate of GPs with total fluence exceeding 50000 Jy~ms. Identically to  Fig. \ref{fig:rate} positions of glitches at MJD 55875.5, 57839.92, and 58237.357 are shown by black dash-dotted vertical lines. By far the strongest glitch at MJD 58064.55 is shown by a vertical bold line.}
\label{fig:rate_7200Jyms}
\end{figure}

The evolution of observed rate of GPs is presented in Figs. \ref{fig:rate} and \ref{fig:rate_7200Jyms}. In Fig. \ref{fig:rate} rate of all detected strong individual pulses is demonstrated (at $> 4 \sigma_{noise}$ level, that puts the lower edge of further investigated data set to about 120 Jy level. Taking into account the assumption that the Crab pulsar radio emission in the main pulse and interpulse consists entirely of giant radio pulses \cite{Popov2006-1}, that would effectively exclude faint 'regular' pulses, which form main pulse precursor, from further consideration). An example of a Crab giant pulse is  presented in Fig.~\ref{fig:GP}. Fig.~\ref{fig:double_GP} shows an example of sequence of two strong GPs in one session separated by 5 periods of the pulsar.

A very sharp peak around epoch MJD$\sim$58100 can be easily seen. It would be tempting to connect it to a very strong glitch which took place at approximately that time (MJD 58064), however it could be just  a coincidence and should be discussed  in more details.  First, not every glitch in the studied timespan was accompanied by a corresponding increase in the production rate, although it should be noted that it can be explained by  low  amplitudes of other glitches --  apart from one at MJD 55875 all of them were weaker by more than two orders of magnitude. Second, there was  strong  'orphan' increase of activity at MJD 56700-57200 which was unaccompanied by any glitch. On the other hand, there are some indications on non-standard rotational behaviour of the Crab pulsar during this epoch \cite{Vivekanand2017}. Thus it can be said that  there are only  weak evidences of correlation between glitch activity and GP production rate at the very best except the fact that the maximum of GPs is delayed by $\sim$150~days with respect to the glitch epoch  MJD 58064.

Still non-uniform  rate of detection  shall be explained. The most obvious explanation would be instrumental effects produced by  the telescope, but it is rather difficult  to reconcile with our absolute calibration technique  and thorough visual checks of candidate pulses. Additionally, the variations in  effective area of the telescope, that if unaccounted for could potentially affect the signal-to-noise ratio and therefore the number of detected strong pulses, were controlled using a set of 
several calibrating sources  and showed no anomalies which correlate with the rate of detections.

\begin{figure}
\begin{center}
\includegraphics[scale=0.6]{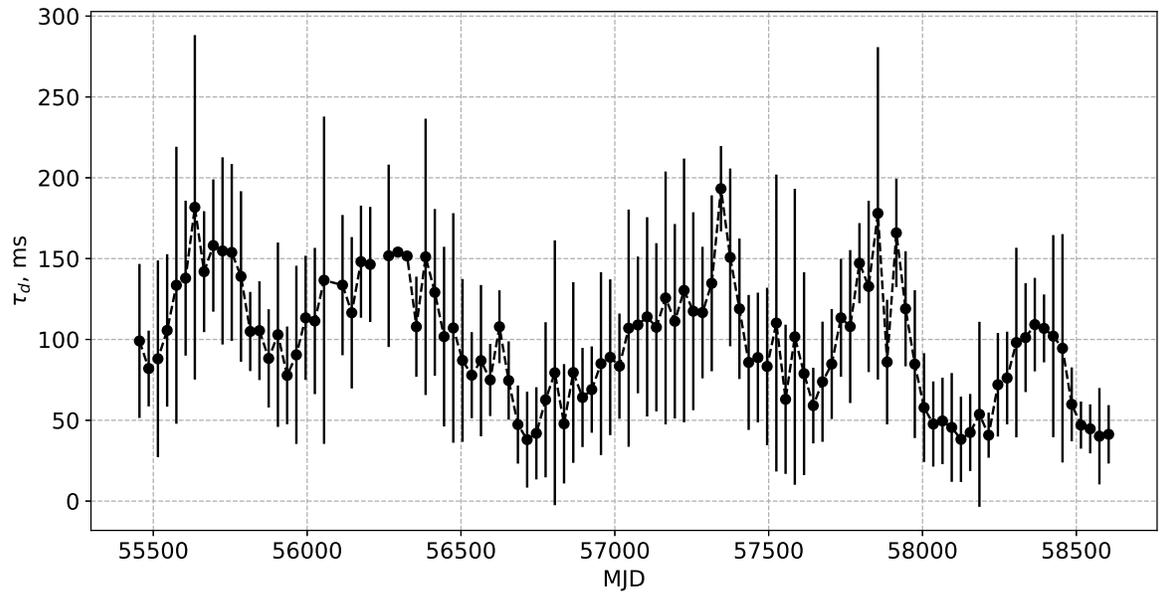}
\end{center}\caption{Scattering time $\tau_d$ averaged in 30-day long bins. Error bars represent  standard deviation of scattering times in respective bins. }
\label{fig:width}
\end{figure}

\begin{figure}
\begin{center}
\includegraphics[scale=0.6]{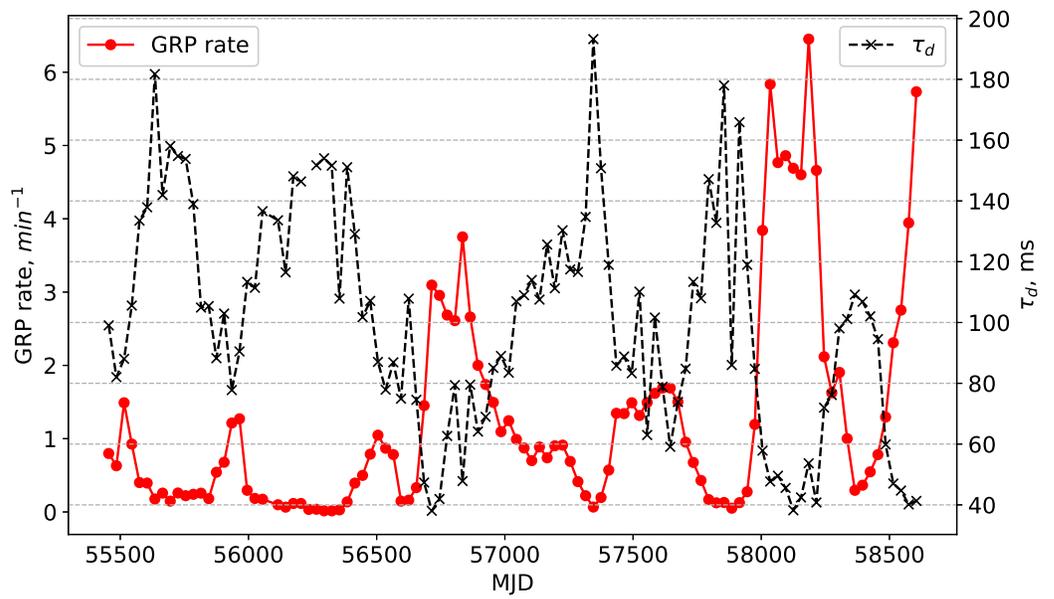}
\end{center}
\caption{The average scattering time $\tau$ and the rate of GPs with peak flux density larger than $4\sigma_{noise}$. Each point contains 30-day averaged data.}
\label{fig:scatter_rate}
\end{figure}

Propagation effects, especially strong at low frequencies, could also be responsible for the observed anomaly. We used the cut based on the observed peak flux and this value could be easily influenced by the scattering. Indeed, even if the generation rate of the GRPs with some fixed fluence $F$ was absolutely stable, peak flux  $S \propto F/\tau_d$ would be modulated by the changing behavior of  $\tau_d$ (or, equivalently, $W_{50}$) (see Fig.\ref{fig:width}) \cite{Smirnova2009, Losovskii2017}. As it can be seen in Fig. \ref{fig:scatter_rate} there exists a    certain  level of  anticorrelation between the production rate and scattering strength,  the sharpest peak coincides with a strong, although not the strongest one,  decline in pulse widths in MJD 58100-58140 interval  -- our explanation seems to be valid.

\begin{figure}
\begin{center}
\includegraphics[scale=0.6]{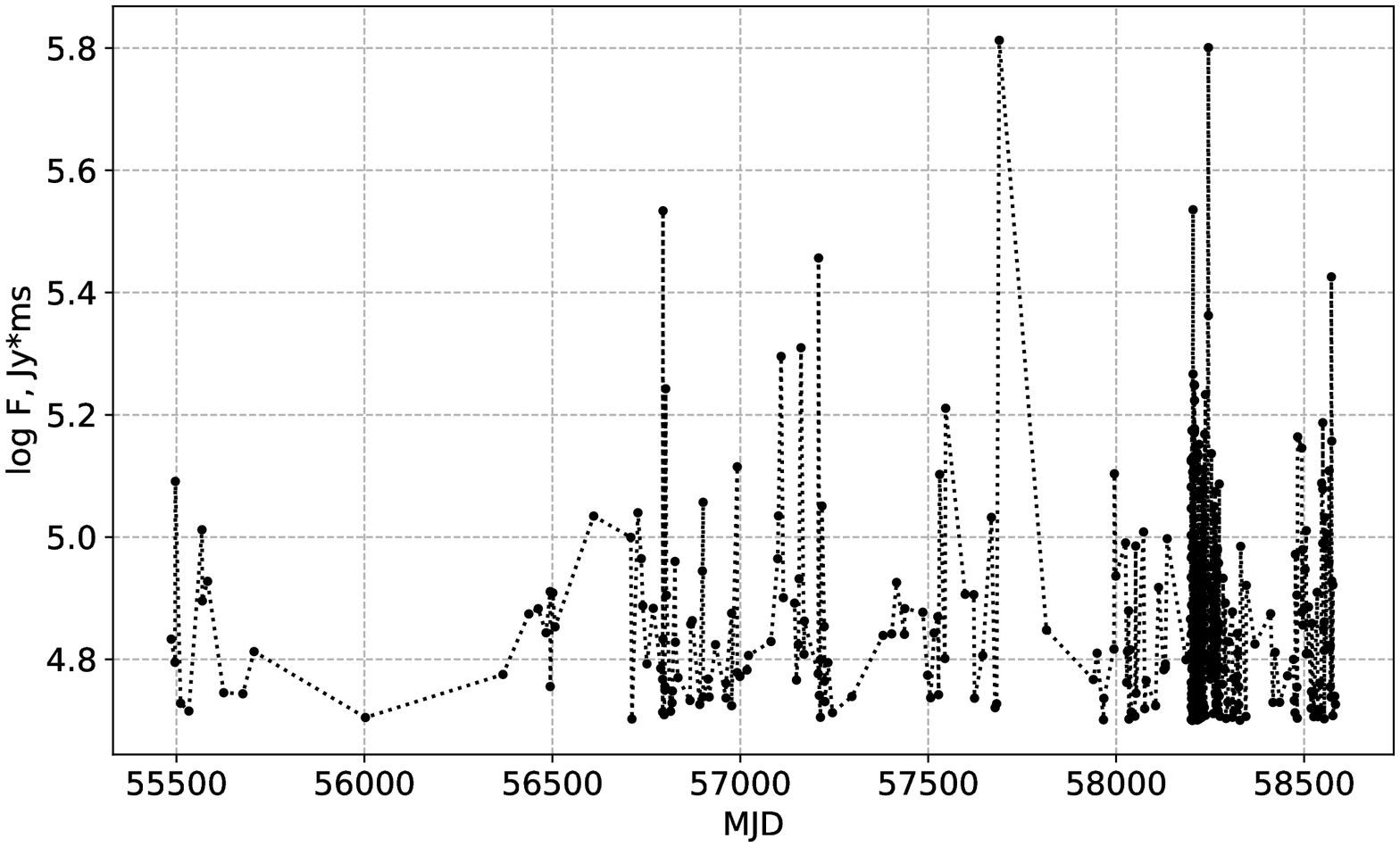}
\end{center}
\caption{Set of the fluences of GPs exceeding 50000 Jy~ms.}
\label{fig:fluence_cut_E}
\end{figure}

 It could be seen that the maximal production rate in  2014-2015 was at MJD 56850 and was higher than the rate during the outburst at MJD 56730. On the contrary, the scattering time in the latter epoch, 20 ms  was much lower than the 30 ms value around MJD 56850. Next,  over some threshold in fluence even the strongest scattering could not prevent GP from entering our set, and thus we would expect some rather uniform distribution of such pulses. In Fig. \ref{fig:rate_7200Jyms} we present the distribution of GRPs with $F>50000~\mathrm{Jy~ms}$. These pulses must be detected  if  scattering times  do not
exceed  150 ms, which is true for a larger part of our observations (see Fig. \ref{fig:width}). This additional fluence cut makes our analysis more robust to uncertainties caused by the scattering.  The peak around MJD 58200 is even more pronounced now (Fig. \ref{fig:rate_7200Jyms}) and it is quite difficult to explain with an assumption that the intrinsic rate of GRP generation is stable. Also the minor activity in MJD 56668-57252 still could be detected. This non-uniform behaviour of pulses  with the largest fluences can also be seen in Fig. \ref{fig:fluence_cut_E}. 

\begin{figure}
\begin{center}
\includegraphics[scale=0.6]{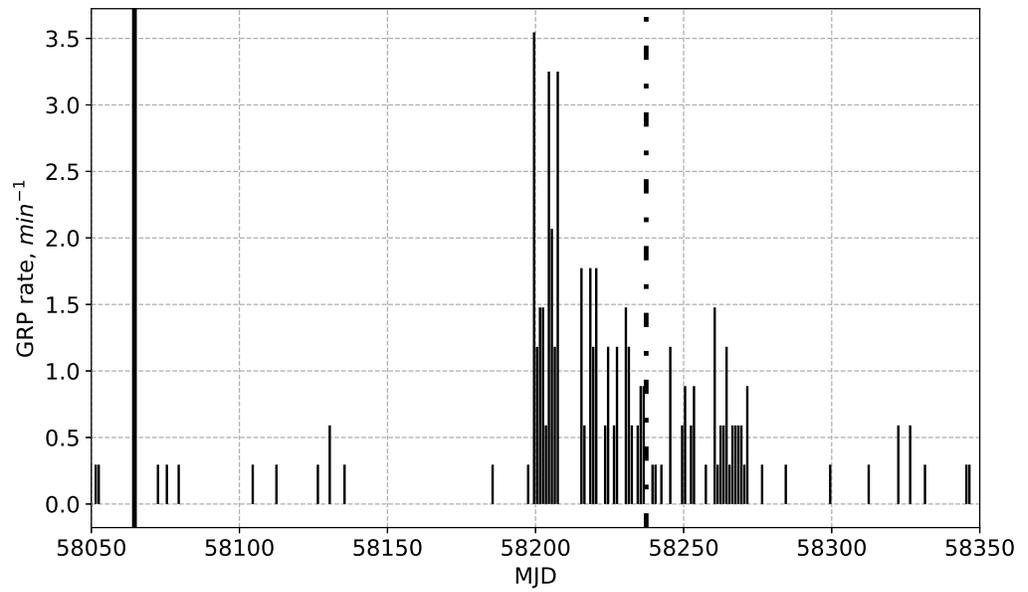}
\end{center}
\caption{The rate of GPs with total fluence exceeding 50000 Jy~ms in  1-day long bins. Identically to  Fig. \ref{fig:rate} position of glitch at MJD 58237.357 is shown by black dash-dotted vertical lines. By far the strongest glitch at MJD 58064.55 is shown by a vertical bold sold line.}
\label{fig:rate_1dayres}
\end{figure}

To investigate more thoroughly region after strongest glitch at MJD 58064.55 the rate of GPs emission with daily time resolution was calculated. As we can see in Fig.\ref{fig:rate_1dayres} (absence of event near MJD 58210 is due to lack of observational data) a strong peak near MJD 58200 shows a sharp rise (from zero to ~3.5 $\mr{min}^{-1}$ in less than 1 day that is minimal time resolution of our data) and exponential decay at about 100 days time interval. Such a behavior is typical for relaxation process in dynamical system and hardly could be explained by propagation effect in the nebula.

Taking into account typical properties of dense clouds that can contribute to quick variations in scattering and DM \cite{Kuzmin2008} ($n\sim 10^{3}~\mathrm{cm^{-3}}$, $T \sim 10^{4}~$K, $d \sim 10^{14}~$cm) it can be estimated that optical depth to free-free absorption even at 100 MHz is much smaller than unity, so these modulations could hardly be effects of obscuration by dense plasma clouds. 

Intriguingly, a  strong jump in DM (and scattering) in 1997 that was  later  explained as an effect of interaction with a cloud in the nebula  also coincided with a weak glitch  MJD 50812 \cite{Backer2000,Lyne2001}.

\section{Conclusions}
\label{sec:conclusion}
After analysis  of almost nine years of everyday observations of the Crab pulsar we constructed a  large sample of bright GPs. The distribution of detected pulses is highly non-uniform and its properties could not be fully explained by the effects of pulse  scattering in the nebula meaning that  the underlying  distribution is intrinsically non-uniform -- rate of GP generation fluctuates with time. 

After removing possible influence of scattering on our statistics by cutting off GPs with a fluence less than 50000 Jy ms, we still could found the highest peak of the rate of GPs emission near MJD 58200. The GPs' rate rises sharply from zero to ~3.5 $\mr{min}^{-1}$ in time less than 1 day and exponentially decay during about 100 days

Finally, we still were not able to refute a hypothesis that the highest peak in the generation rate is somehow connected with the strongest glitch which precedes it by ~ 140 days.


\section*{Acknowledgements}
AK and MP acknowledge the  support by  the Foundation for the Advancement of Theoretical
Physics and Mathematics <<BASIS>> grant 18-1-2-51-1.
The development of identification of GPs with methods of machine learning was  supported by the Russian Science Foundation grant 17-72-20291.
MP acknowledges support of 
Leading  Science School MSU (Physics of Stars, Relativistic Compact Objects and Galaxies). This research has made use of NASA's  Astrophysics Data System.

\bibliographystyle{unsrt}


\begin{figure}
\begin{center}
\includegraphics[scale=0.6, angle=0]{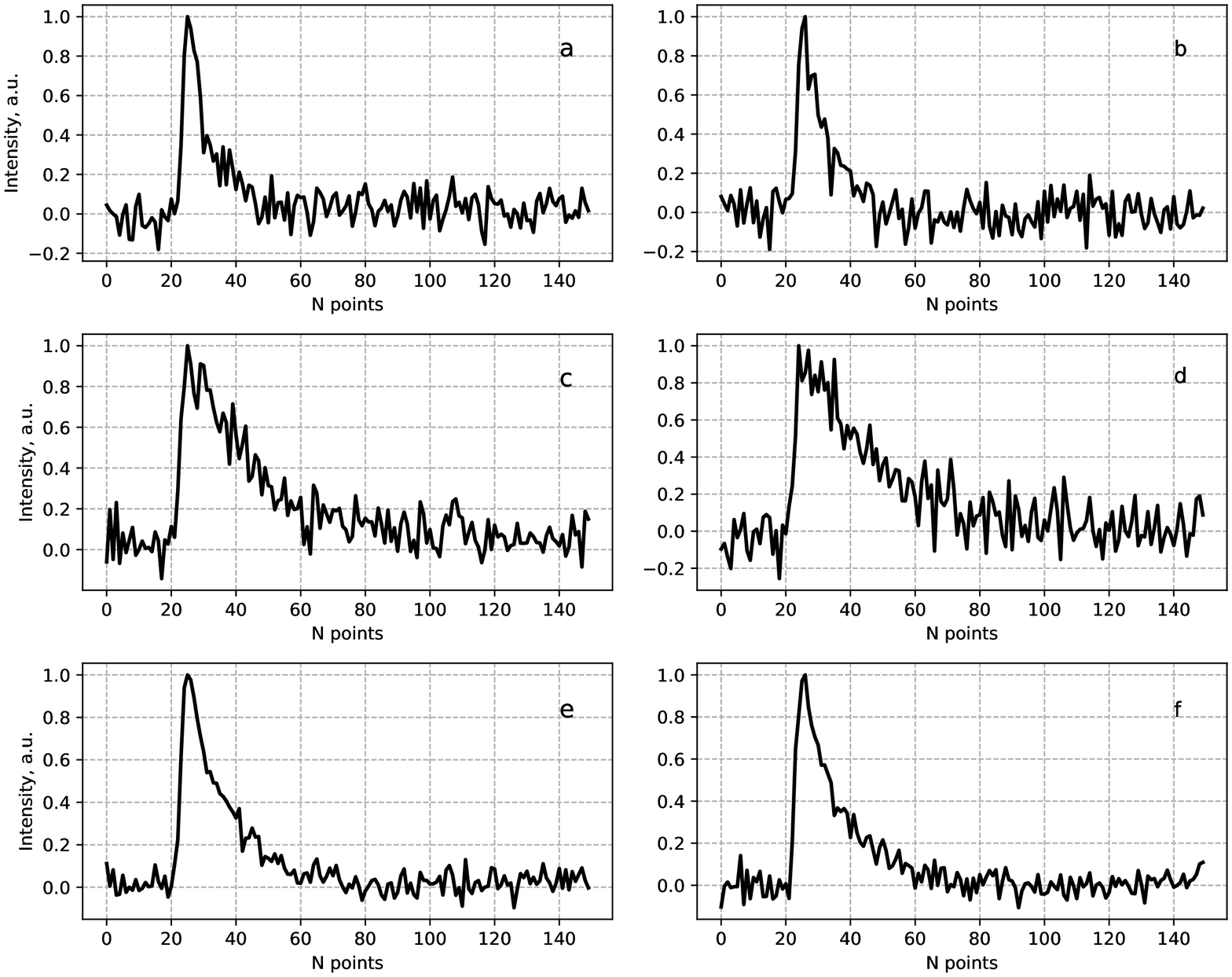}
\end{center}
\caption{Examples of real pulses from B0531+21 (a, c, e) and their corresponding generated fake pulses (b, d, f).}
\label{fig:pulse_generator}
\end{figure}  

\begin{figure}
\begin{center}
\includegraphics[scale=0.6, angle=0]{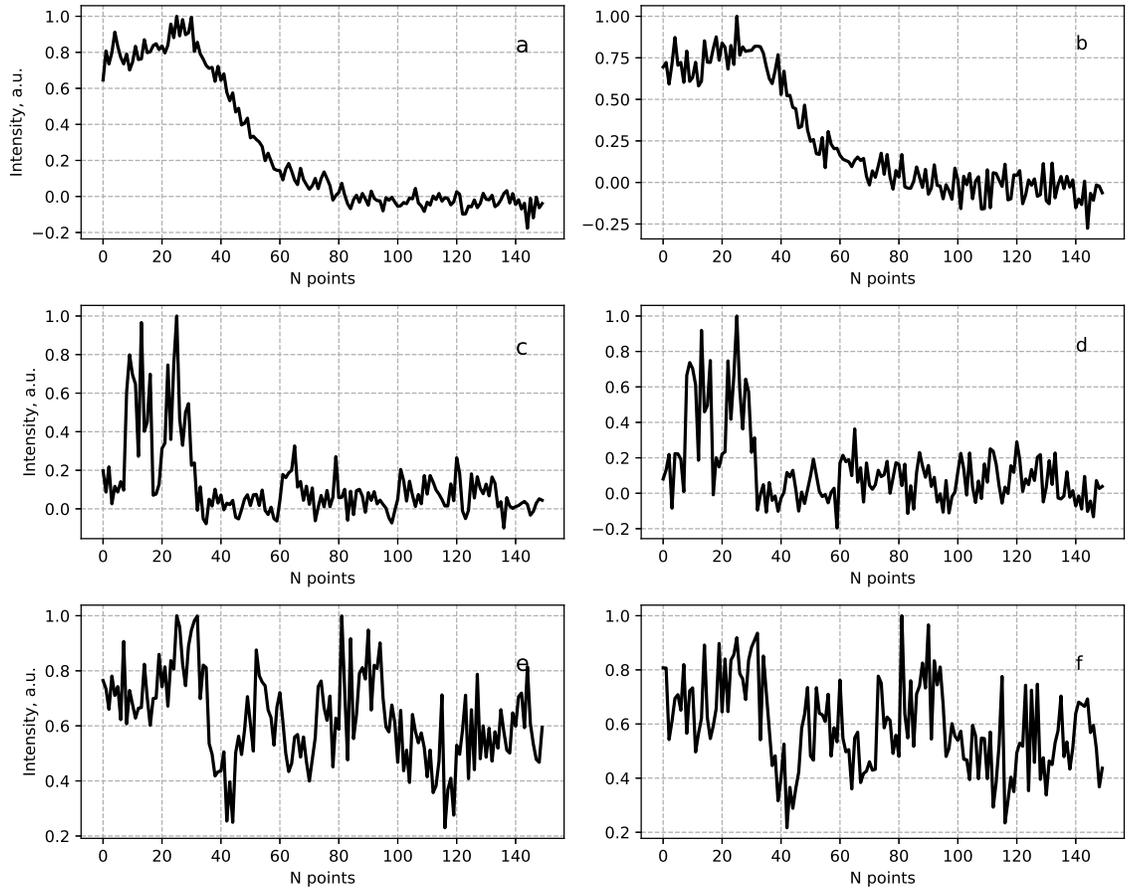}
\end{center}
\caption{Examples of detected noises during B0531+21 observations (a, c, e) and their corresponding generated fake noises (b, d, f).}
\label{fig:noise_generator}
\end{figure}

\begin{figure}
\begin{center}
\includegraphics[scale=0.8, angle=0]{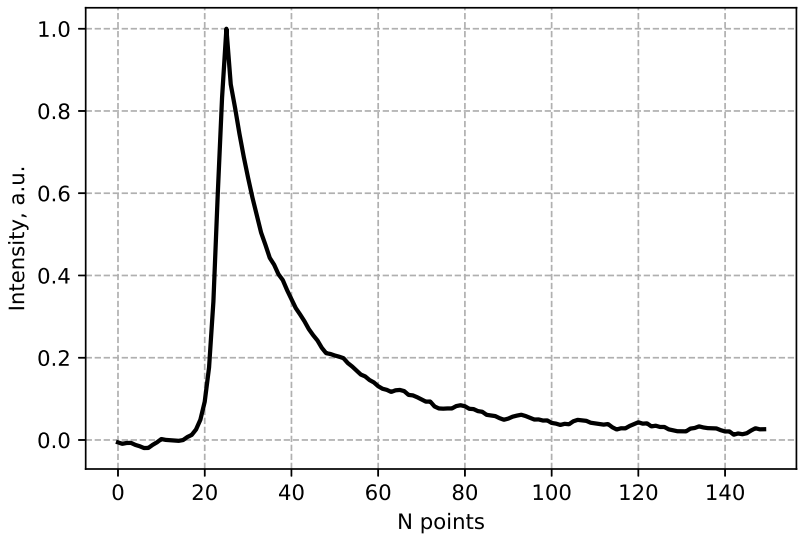}
\end{center}
\caption{The averaged pattern of B0531+21 giant pulses}
\label{fig:total_pattern}
\end{figure} 

\begin{figure}
\begin{center}
\includegraphics[width=13.0 cm,angle=0]{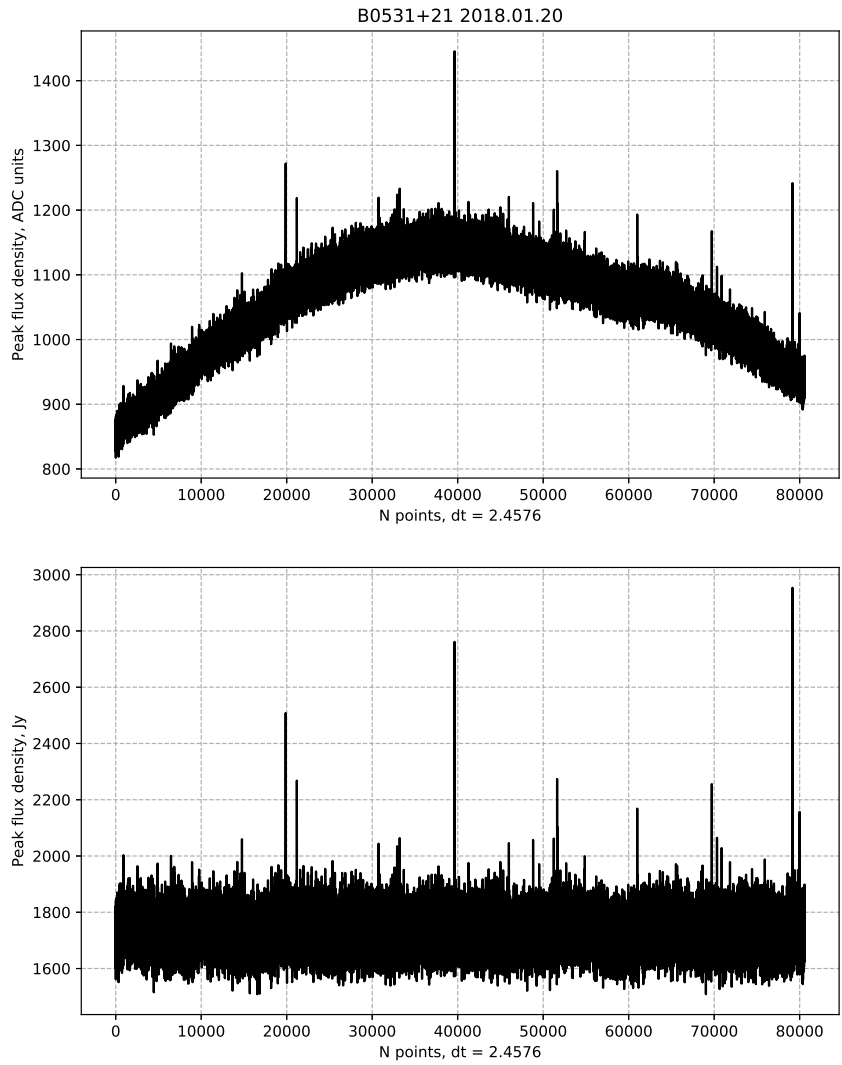}
\end{center}
\caption{Upper panel: an example of  typical Crab transit. Lower panel: the same observation corrected for the beam factor and influence of ionospheric effects. The baseline is normalized to 1720 Jy level. Several strong pulses could be easily seen.} 
\label{fig:diag}
\end{figure}

\textit{.}


\pagebreak
\appendix
\section{Algorithm of GP selection} \label{sec:A}
\label{app-ML}
For pulse recognition we employed methods of machine learning, well suited for aims of extraction of signals with certain shape from large amount of noisy data. First we organized so-called training set: pulses with peak flux density more than 10$\sigma$ were selected and manually divided into two groups: pulses from the pulsar and noises of various nature. This stringent cut allowed to select  characteristic representatives of corresponding classes.  The obtained data set contained only 2271 events. This limited amount was not sufficiently large for the effective training of the classifier, so we needed to generate additional artificial signals in order to augment our training set.   These signals were generated using   real events as their prototypes. Each real pulse was fitted by the function:

\begin{equation}
 f(t) = \int f_{0}(t)s(t-\xi)\xi,
 \label{pulse_formula:ref}
\end{equation}

where:

\begin{equation}
 f_{0}(t) = ae^{-\frac{(t-b)^{2}}{2\sigma ^{2}}}
 \label{gauss_formula:ref}
\end{equation}

and

\begin{equation}
 s(t) = \left\{\begin{matrix}
0 & ,if & t<0\\ 
e^{t/\tau } & ,if & t\geqslant 0.
\end{matrix}\right.
 \label{gauss_formula:ref2}
\end{equation}

where $\tau$ is the scattering time.

After fitting operation, these parameters were used to generate artificil pulse. After that, random noise  with parameters taken from noise  parameters of real pulses (form 0 to 15 points of a record) was added to the artificial  pulse (see Fig. \ref{fig:pulse_generator}). The process of generating fake noises wasn't so complex.  The random noise was just added to the original noise (see Fig. \ref{fig:noise_generator}).
Totally, a set of 20,247 events was created.

The classifier was based on random forests method. We estimated the efficiency of the classifier, using our validation set (567 out of 2271 original, N1 pulses, N2 noises) and it reached  97,8\%.  

Afterwards, the classifier was used for selecting pulses with peak flux density more than 4$\sigma_{noise}$. For all pulses the following  parameters were: time of arrival, peak flux density, $W_{50}$ and $W_{10}$ --  width of pulse at $50\%$ and $10\%$ of peak, correspondingly, and fluence. The last parameter was calculated as an integral of a fitting function (Eq. \ref{pulse_formula:ref}).
At  our next step,  we calculated Pearson correlation coefficient between pulse and averaged pattern of giant pulses, which was  constructed from the most typical detected Crab pulses (see Fig. \ref{fig:total_pattern}). Only pulses with $r>0.5$ were used for further analysis.
Finally the candidate
pulses were visually rechecked.
\end{document}